\documentclass{article}
\usepackage[T2A]{fontenc}
\usepackage[russian,english]{babel}

\usepackage{authblk}
\usepackage{graphicx}
\usepackage{wrapfig}
\usepackage{url}
\usepackage{hyperref}
\usepackage{geometry}

\hypersetup{colorlinks=false,frenchlinks=false}

\title{Differential method for measuring periodic magnetic structure}
\author{Grigory Yakopov\thanks{grigory.yakopov@desy.de} }
\author{Pavel Vagin\thanks{pavel.vagin@desy.de} }
\affil{Deutsches Elektronen-Synchrotron DESY, Hamburg, Germany}

\date{}

\begin{document}
\maketitle

\begin{abstract}
The improvement of accelerator parameters requires the transition to superconducting undulator technologies since SCUs allow higher magnetic fields compared to classical planar undulators with the same gap~\cite{bib:casalbuoni}. The primary method for characterizing magnetic structures is to measure the magnetic field distribution along the undulator axis with a Hall probe.
Special test rigs based on vertical cryostats with liquid helium are used to parameterize superconducting magnetic structures (undulator coils). However, during the probe immersion and cooling to cryogenic temperatures, its linear dimensions change, which needs to be taken into account by applying complex thermodynamic models since the position encoder of the probe is located on the cryostat flange at room temperature. To eliminate this effect, it is proposed to equip the probe with two Hall sensors spaced apart half a period of the measured structure.

%Examples are CASPER~1(KIT)\cite{bib:casper} and SUNDAE~1(EuXFEL)\cite{bib:sundae}.

\end{abstract}

\section*{Introduction}

\begin{wrapfigure}{r}{0.3\textwidth}
  \begin{center}
    \includegraphics[width=\linewidth]{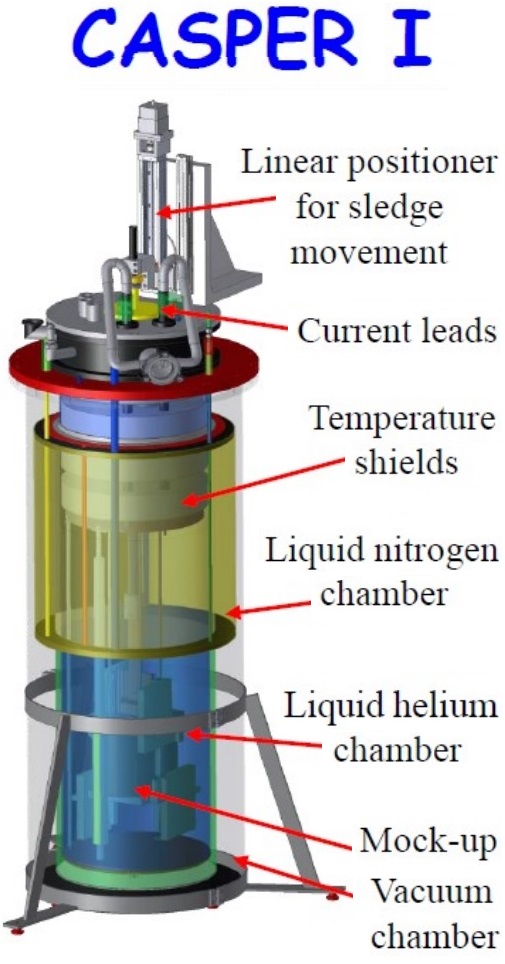}
  \end{center}
\caption{Casper-I}
\label{fig:casper}
\end{wrapfigure}

The differential measurement method is widely used in precision measurements, especially in interferometric control methods\cite{bib:interferometry}. In our study, we show that this method can be used to measure the magnetic field distribution along the axis of a superconducting undulator coil.
The experimental setup CASPER~1 of the Karlsruhe Institute of Technology (KIT) (fig.~\ref{fig:casper}) consists of the measured (parameterized) magnetic structure positioned vertically in a cryostat containing liquid helium, and a vertical probe with a Hall sensor that can be moved along the structure\cite{bib:casper}. For the same purposes, a setup called SUNDAE~1 is currently being developed for the European XFEL\cite{bib:sundae}.

%\clearpage
\section*{Suppression of the main harmonic}
Rotating coil is a common method for the measurements of accelerator multipole magnetic lenses. Bucking coils are used to suppress the main harmonic and measure low-level signals of higher harmonics. Bucking coils are connected in series opposite to the measurement coil, and with different radius and numbers of turns could fully suppress the main harmonic, or any other particular harmonic, while being sensitive to the remaining harmonics\cite{bib:buckcoil}.

A similar approach could be used to suppress the main harmonic of the undulator field and measure only residual errors. This would make the measurement system less sensitive to position errors as there would be no large signal from the main harmonic with large gradients at zero crossings. Also, field integrals are not affected by filtering of the main harmonic, as long as the filter has 0dB response at zero frequency.
Suppression of the main harmonic could be done by placing two Hall sensors at a distance of exactly half of the undulator period or using a search coil with a length of undulator period, $\lambda$-coil.
\clearpage
\begin{wrapfigure}[15]{tr}{0.35\textwidth}
%\begin{figure}
\begin{center}
\vspace{14.5mm}
\includegraphics[width=\linewidth]{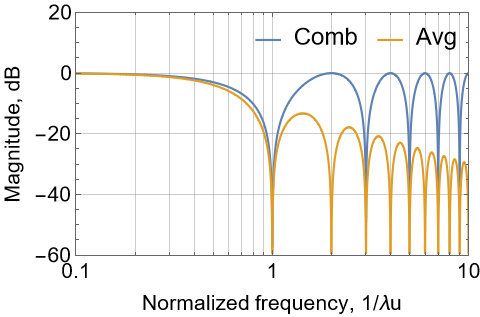}
\end{center}
\caption{Frequency response of comb and moving average filters}
\label{fig:filter}
%\end{figure}
\end{wrapfigure}
\
\subsection*{Two Hall sensors displaced by half period}
Two Hall sensors, displaced by half of the undulator period, form a comb filter\cite{bib:combfilter}:
\mbox{$Y(s) = X(s) + \alpha X(s-\lambda_u/2)$}, where $\lambda_u$ is undulator period and $\alpha=+1$ in case of measuring sum signal of two Hall probes. Such a comb filter suppresses the main harmonic of the sinusoidal undulator field and also its odd harmonics (fig.~\ref{fig:filter}).

\subsection*{$\lambda$-coil}
Search coil with the length of undulator period acts as moving average filter\cite{bib:avgfilter} applied to the measured undulator field. It has a similar to comb filter frequency response, except being a full period coil it suppresses all the higher harmonics, not only odd ones, and because of integrating field over coil volume (undulator period) it has an additional 20dB/decade ($\sim1/\omega$) suppression of higher frequencies (fig.~\ref{fig:filter}).

\section*{Position errors}
\begin{figure}[h]
    \centering
    \includegraphics[width=\textwidth]{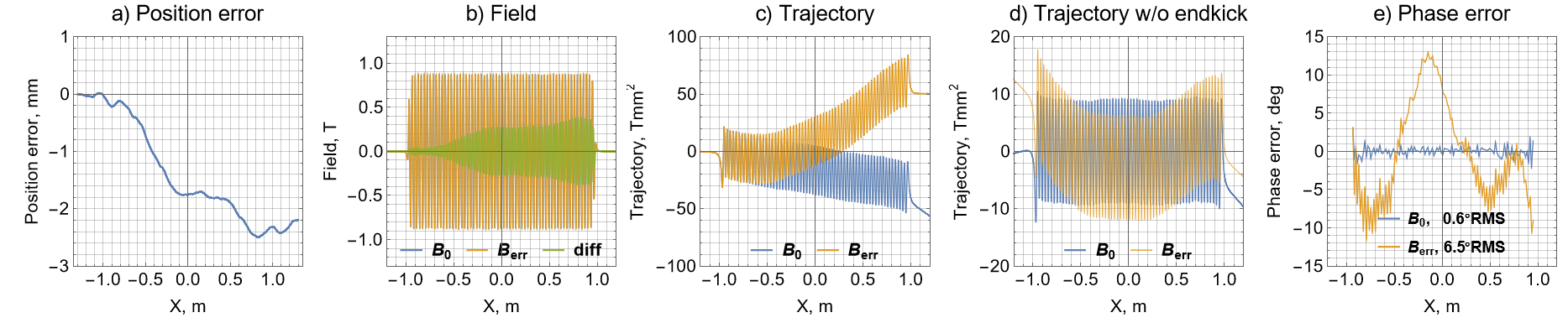}
    \caption{Measured magnetic field (b) is resampled with additional position error (a), causing additional errors in trajectory (c, d), and phase error(e)}
    \label{fig:posErr}
\end{figure}

Figure \ref{fig:posErr} shows a measured magnetic field of a real 2m undulator with a 31.4mm period, which is then resampled with additional random position error (fig \ref{fig:posErr}a) to simulate errors during measurement. This causes changes in the calculated electron trajectory and degradation of the phase error from less than 1 degree RMS of the tuned undulator without position error to more than 6 degrees RMS when "measured" with additional position error.

\subsection*{Filtering field data}
\begin{figure}[h]
    \centering
    \includegraphics[width=0.85\textwidth]{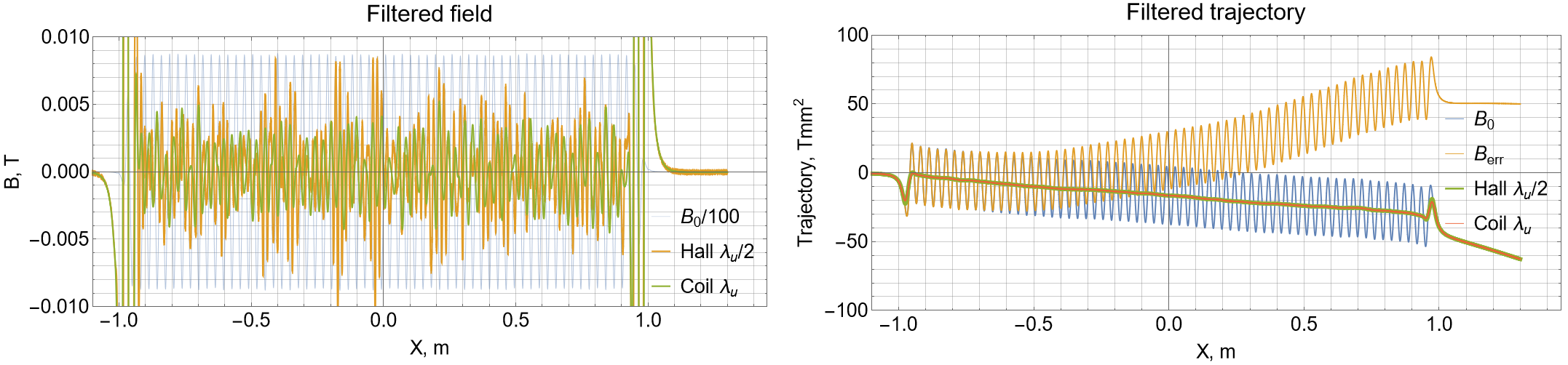}
    \caption{Comb filter and moving average filter applied to measured undulator magnetic field data. Trajectories calculated from the field measured with a single Hall probe, with($B_{err}$) and without($B_0$) position error, an average of two Hall probes spaced by $\lambda_u/2$ (comb filter) and $\lambda$-coil (moving average filter). Filtering with two Hall sensors or  $\lambda$-coil suppresses main harmonic and large field gradients, so additional position errors do not affect the second field integral (trajectory) calculated from the filtered data.}
    \label{fig:Bfilt}
\end{figure}

Applying a comb filter (measuring the average of two half-period spaced Hall probes) or moving average filter (measuring with $\lambda$-coil) to the field data suppresses the main harmonic. Figure \ref{fig:Bfilt} left plot shows residual field errors after filtering (difference from ideal sinusoidal main harmonic with harmonics).

Both filters do have a 0dB response at a frequency of 0Hz thus they both do not affect field integrals. And as filtering of the main harmonic eliminates large field gradients,  position errors without large field gradients do not contribute to the field errors. Thus electron trajectory calculated from filtered field data with position error is exactly the same as the trajectory calculated from field data without position error, except for suppression of the oscillation of the main harmonic (fig. \ref{fig:Bfilt} right plot).

\subsection*{Reconstructing position error from measured field data}
Measured data $B_{err}$ obtained by a single sensor could be compared to the filtered data obtained by two Hall sensors or $\lambda$-coil. Single sensor data is affected by position error, while measurement which is "filtered" with two Hall sensors or $\lambda_u$-coil is insensitive to position error. So the difference of the fields measured with two sensors (prefiltered data) and fields measured with a single sensor but filtered afterward (postfiltered) would contain information about position error and could be used to recover the actual position error. The difference between "prefiltered" and "postfiltered" data is multiplied by a field in order to consider the sign and integrated. The result is proportional to the position error.

\begin{figure}[h]
    \centering
    \includegraphics[width=0.75\textwidth]{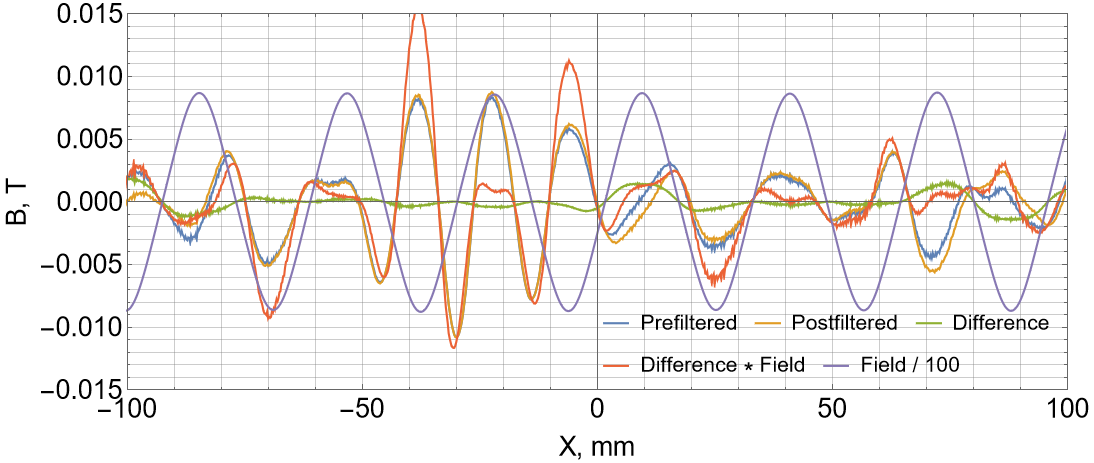}
    \caption{Magnetic field measured with two Hall probes (comb filtered) vs magnetic field of single Hall probe filtered afterward, their difference and difference multiplied by field}
    \label{fig:posErrRecovery}
\end{figure}

\begin{figure}[h]
    \centering
    \includegraphics[width=\textwidth]{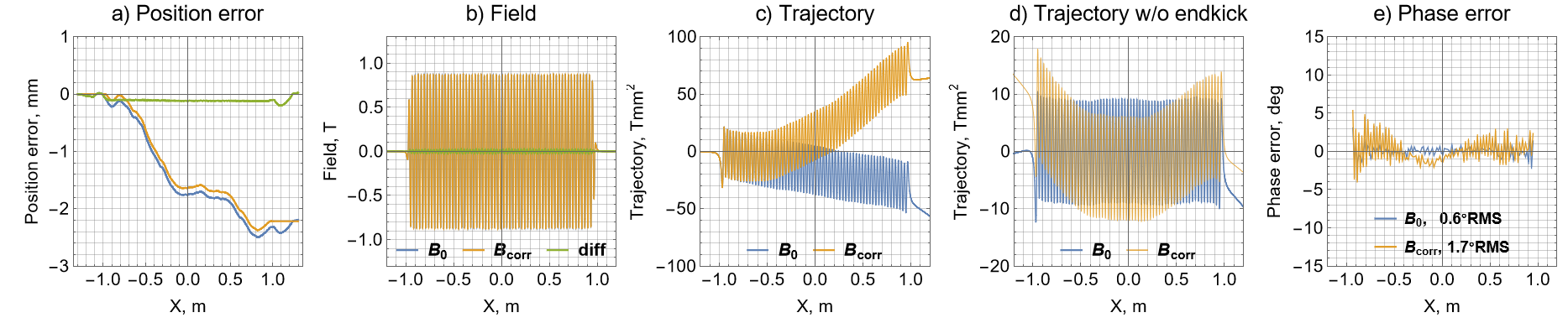}
    \caption{Magnetic field (b) measured with position error (a) is resampled at proper positions, correcting field difference(b) and phase error(e), but not trajectory errors (c, d), that caused by high-frequency position jitter}
    \label{fig:posErrCorr}
\end{figure}

There is some difference in recovered position error (fig. \ref{fig:posErrCorr}a), as outside of the structure there is no field amplitude and thus no information to recover, but except this small offset, which is accumulated outside, the position error inside the structure could be reconstructed using comb filtering with two Hall sensors or $\lambda$-coil.

Having the actual position error, recovered from measured data, it is possible to resample single sensor data $B_{err}$ at proper positions to compensate for the position error. However, despite two sensors spaced with a half period could suppress the main harmonic and thus not sensitive to position error, reconstruction of the position error does not work perfectly well at the field maximums, where the field gradient became zero for both sensors simultaneously. At these locations, this method of reconstruction becomes not sensitive to the position error, as both sensors show the same maximum value and some position error in zero gradients around the maximum would not affect the sensor readings much. Still could be used to reconstruct long-range errors caused by "slow" changes of position because of for example temperature variation. However high-frequency jitter of position would introduce additional errors that could not be recovered by this method. High-frequency position errors are mostly responsible for trajectory errors and figure~\ref{fig:posErrCorr} in comparison with figure~\ref{fig:posErr} shows that correction of a position error, recovered from field data, has improved the field difference and phase errors, but did not correct the trajectory errors. And remaining phase error is caused by non-corrected trajectory errors. If the electron beam while traveling through the undulator got some trajectory kicks, it also would emit radiation at varying angles, which slightly changes the wavelength ($\sim1+\theta^2 \gamma^2$) and thus accumulates additional phase error.

However we also do have filtered field data measured with two Hall probes, which results in a proper trajectory without error, but with main harmonic suppressed (fig. \ref{fig:Bfilt}). The difference between "prefiltered" and "postfiltered" field data, could be added to measured field $B_{err}$ and correct also for the trajectory and remaining phase error.

\begin{figure}[h]
    \centering
    \includegraphics[width=\textwidth]{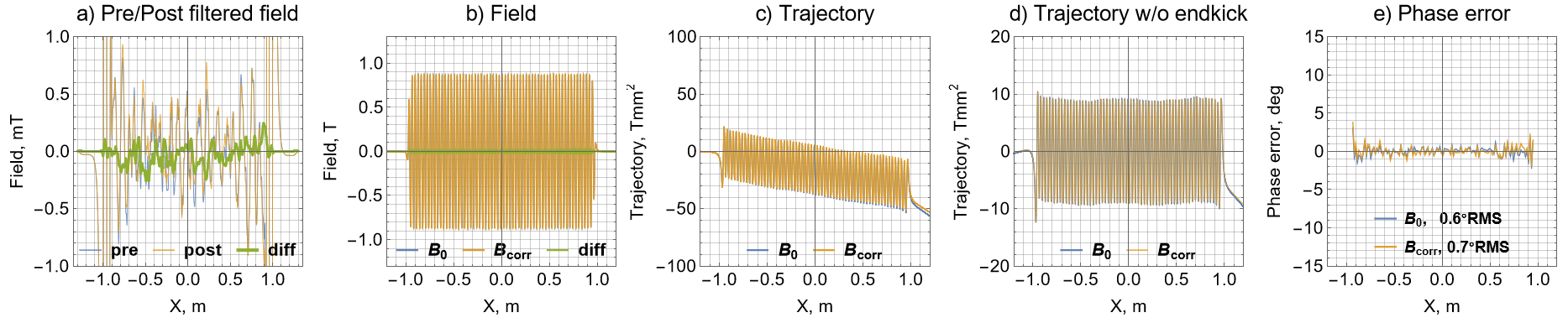}
    \caption{Additional field correction, calculated from the difference of fields, filtered during measurement with two Hall probes and fields of single probe but filtered afterward, results in correction of the trajectory error and residual errors of the phase error, with negligible difference to the initial measurements of trajectory and phase error without additional position error}
    \label{fig:posErrCorrTraj}
\end{figure}

Two Hall sensors displaced by a quarter of the undulator period would not get in zero field gradient simultaneously and could act as sin/cos encoders using a sinusoidal undulator magnetic field. A similar approach with comparing values of two vertically displaced sensors in a nonuniform undulator magnetic field with known $B(y) \sim cosh(y)$ vertical dependence is used to determine vertical position error and recalculate actual on-axis field value\cite{bib:vasserman}.

%\iffalse
\section*{Using undulator magnetic field as quadrature sin/cos encoder}
If two Hall sensors are placed at a distance of $\lambda_u/4$, their readings could be used as a sin/cos linear encoder to get the sensor position from the magnetic field data itself. However, one needs to know first the exact field profile to make a position correction to compensate for actual field errors. And actual field profile is measured with the same system and thus contains position errors causing distortions of the measured field profile. 

Using magnetic field as a linear encoder directly by using arc-tangent of two field values sampled at points $x0\pm\lambda_u/8$ results in position errors caused by errors in the magnetic field, and non-sinusoidal field profile as it contains also higher harmonics. 

\begin{figure}[h]
    \centering
    \includegraphics[width=\textwidth]{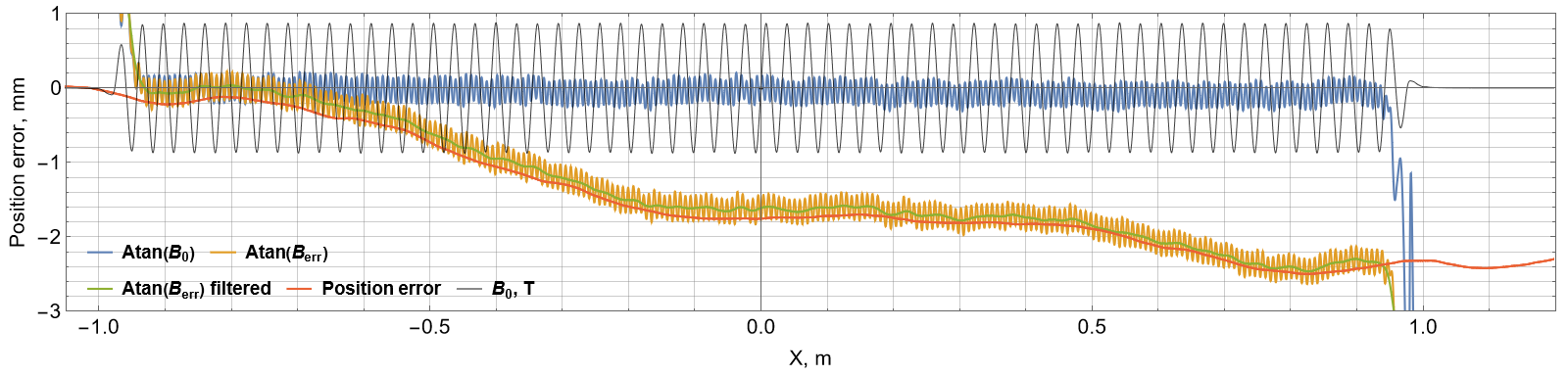}
    \caption{Measured magnetic field of the undulator, and position error derived from the magnetic field as $X = ArcTan(B_{cos}, B_{sin})$, where $B_{cos} = B(x-\lambda_u/8)$ and $B_{sin} = B(x+\lambda_u/8)$ - magnetic field sampled with two sensors displaced by a quarter of undulator period. Additional position errors could be recovered from field data, but a nonideal undulator field with higher harmonics also adds additional position error oscillation that needs to be filtered out.}
    \label{fig:arctanB}
\end{figure}

A naive approach to correct for position error would be to calculate position directly as the arctangent of two sensors (sin/cos) and then filter out the position oscillation caused by higher harmonics of the undulator field (fig. \ref{fig:arctanB}). Like the previous position error recovery from magnetic data, based on two half-period displaced sensors, this could correct for long-term drifts of position errors, but not high-frequency ones. It would correct for phase error but trajectory errors would remain uncorrected and, unlike correction with half-period spacing, there is no additional filtered data which is insensitive to position error to correct trajectory errors.

However the difference between data from one sensor and data from another sensor, but interpolated and resampled at the position of the first sensor could be used to distinguish field errors from sensor position errors and improve recovery of the position error from magnetic field data.

%\fi

\section*{Conclusion}
The proposed method with two-spaced sensors allows to eliminate influence of the position error on the measured undulator field integrals(trajectory) and detect undulator field errors even when accurate longitudinal positioning is not possible. And also to reconstruct actual position error from the field data, using the distance between the probe as a reference, and then to compensate for this error by resampling interpolated measured field data at the proper positions.
%The proposed method allows the measurement of the magnetic field distribution along the axis of the superconducting undulator coil with a resolution of the order of a few milli-tesla and a spatial resolution of the order of a few micrometers, ignoring the influence of the rod on the measurement results.

\end{document}